\documentclass[%
10pt,
reprint,
nobibnotes,
 amsmath,amssymb,
 aps,
prl,
longbibliography
]
{revtex4-1}
\usepackage{graphicx,color}
\usepackage{bm}
\usepackage{wrapfig}
\usepackage{relsize}
\usepackage{amssymb}
\usepackage{upgreek}
\newcommand{\pd}{\partial}
\def\d{\mathrm{d}}

\newcommand{\xibar}{\bar{\xi}}
\newcommand{\etabar}{\bar{\eta}}
\newcommand{\refLvsSQproof}{S.I}
\newcommand{\refrecLproof}{S.II}
\newcommand{\refsecLvsSp}{S.III}
\newcommand{\refproofQPvsL}{S.IV}
\newcommand{\refproofLvsQP}{S.V}
\newcommand{\refproofQPbarMain}{S.VI}
\newcommand{\refintLnmProof}{S.VII}
\newcommand{\refSQsourceproof}{S.VIII}

\newcommand{\refLvsS}{3}
\newcommand{\refLvsSQ}{4}
\newcommand{\refEqnQoffset}{12}
\newcommand{\refLrec}{16}
\newcommand{\refLisolated}{19}
\newcommand{\refLvsSp}{20}
\newcommand{\refQPbarvsL}{22}
\newcommand{\refLvsQPbar}{23}
\newcommand{\refQPbarMain}{24}
\newcommand{\refintQPbar}{26}
\newcommand{\refintLnm}{27}
\newcommand{\refintSQ}{28}

\usepackage{graphicx}

\begin{document}
\title{A new class of solutions to Laplace equation: Regularized multipoles of negative orders}
\author{Matt Majic}
\author{Eric C. Le Ru} \email{eric.leru@vuw.ac.nz}

\affiliation{The MacDiarmid Institute for Advanced Materials and Nanotechnology,
School of Chemical and Physical Sciences, Victoria University of Wellington,
PO Box 600, Wellington 6140, New Zealand}

\date{\today}

\begin{abstract}
We introduce a new class of solutions to Laplace equation, dubbed logopoles, and use them to derive a new relation between solutions in prolate spheroidal and spherical coordinates. The main novelty is that it involves spherical harmonics of the second kind, which have rarely been considered in physical problems because they are singular on the entire $z$ axis. Logopoles, in contrast, have a finite line singularity like solid spheroidal harmonics, but are also closely related to solid spherical harmonics and can be viewed as an extension of the standard multipole ladder toward the negative multipolar orders.
These new solutions may prove a fruitful alternative to either spherical or spheroidal harmonics in physical problems.
\end{abstract}
\maketitle


Laplace equation is the fundamental equation in a wide range of physical problems including astronomy, geophysics, electrostatics, and fluid mechanics \cite{Morse1953,1998Jackson,2008Stacey,2012Pavlis}. It is also strongly related to the even more pervasive Helmholtz equation, which governs wave phenomena for example in acoustic and electromagnetic scattering. The latter link is both mathematical, since the solutions have similarities, and physical, since Laplace equation is the long-wavelength limit of Helmholtz equation.
The solution of Laplace equation using the separation of variable method is a standard textbook problem \cite{Morse1953,1941Stratton}.
In spherical coordinates ($r$,$\theta$,$\phi$), the angular part of the solution consists of spherical harmonics, proportional to $P_n^m(\cos\theta)e^{i m\phi}$, where $n$ is a positive integer called the multipole order, $m$ is an integer satisfying $|m|\le n$, and $P_n^m$ is the associated Legendre function of the first kind. The radial part of the solution is of the form $r^n$ (finite at the origin) or $r^{-n-1}$ (singular at the origin, but regular at infinity).
This results in two types of solutions called solid spherical harmonics (SSHs): the internal SSHs, $r^nP_n^m(\cos\theta)e^{im\phi}$, and the external SSHs $r^{-n-1}P_n^m(\cos\theta)e^{i m\phi}$. 
A similar approach exists in other coordinate systems, and we will focus here on prolate spheroidal coordinates ($\xi$,$\eta$,$\phi$) defined by two focal points.
The angular dependence is similar, of the form $P_n^m(\eta)$, while the radial solution involves Legendre functions
of either the first kind $P_n^m(\xi)$ (finite at the origin), or of the second kind $Q_n^m(\xi)$ (singular at $\xi=1$, segment between the two foci, but regular at infinity). The internal and external prolate spheroidal solid harmonics (PSSHs) $P_n^m(\xi)P_n^m(\eta)e^{i m\phi}$ and $Q_n^m(\xi)P_n^m(\eta)e^{i m\phi}$ are commonly used in problems with spheroids or elongated objects. Other types of solutions have more recently been studied \cite{1953Garabedian,1967Martinek,1975Burstein,1990Harp,1992Martinov}, but none as fundamental and generally applicable as SSHs and PSSHs.
Interestingly, it was shown recently that PSSHs also provide an advantage in problems with spherical geometry \cite{majic2017super}, where the SSHs would have {\it a priori} been better suited. Links between SSHs and PSSHs via series expansions were derived more than a century ago \cite{jeffery1916relations}.

\begin{figure}
\includegraphics[scale=.7,trim={5.5cm 10.8cm 5.5cm 10.8cm},clip]{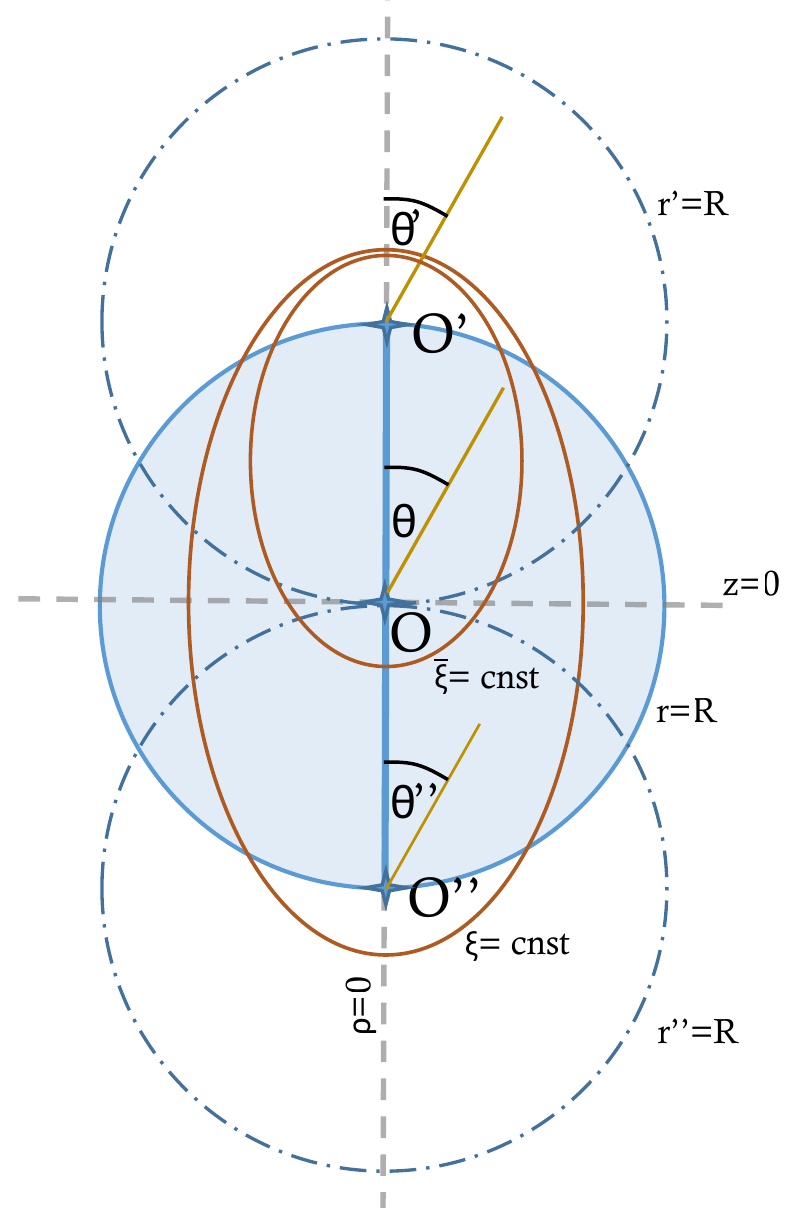}
\caption{Schematic of the centered and offset spherical and prolate spheroidal coordinate systems considered in this work.} \label{CoordinateSystems}
\end{figure}

Both external SSHs and PSSHs have bounded singularities and go to zero at infinity, making them useful for solving problems outside bounded domains. 
Although rarely mentioned or used, there exists equivalent solutions where the angular part takes the form of the associated Legendre functions of the second kind, $Q_n^m(\cos\theta)$. These are normally discarded due to their singularities on the entire $z$-axis, which precludes their application to bounded physical systems. We here present a way around this problem by combining SSHs of the second kind centered at two different origins to remove the singularities at infinity. This approach can be understood simply by considering the lowest order solution $Q_0(\cos\theta)$.
Close to the $z$-axis, it behaves as $Q_0(\cos\theta)\sim \ln(z/\rho)$, where $\rho=\sqrt{x^2+y^2}$, hence the singularity for $\rho=0$. Now let us define two offset coordinate frames with origins O$'$ and O$''$ at $z=R$ and $z=-R$ on the $z$-axis and their associated spherical coordinates $(r',\theta',\phi)$ and $(r'',\theta'',\phi)$, see Fig.~\ref{CoordinateSystems}.
We have for small $\rho$:
\begin{align}
Q_0(\cos\theta') - Q_0(\cos\theta'') &\underset{\rho\rightarrow 0}{\sim} \ln \frac{z-R}{\rho}-\ln \frac{z+R}{\rho}+O(\rho^2) \nonumber\\
&\underset{\rho\rightarrow 0}{\sim} \ln \frac{z-R}{z+R}+ O(\rho^2),
\label{EqnIntro}
\end{align}
which is finite except on the segment between O$'$ and O$''$.
By combining two offset SSHs of the second kind, we have therefore obtained a solution whose singularity is bounded, which makes such combinations
suitable for solving practical problems. Interestingly, the resulting segment singularity is the same as that of the PSSH solutions when O$'$ and O$''$ are chosen as the foci defining the spheroidal coordinates, in fact the solution in \eqref{EqnIntro} is the PSSH of order 0, $Q_0(\xi)$.

In this letter, we explore further this idea. We generalize this simple example to derive a new relationship between the uncommon SSHs of the second kind and the more common PSSHs.  To prove this relation, we introduce a new class of solutions, dubbed logopoles, that are closely related to both spherical and spheroidal harmonics. Like the spheroidal harmonics, they are singular on a bounded line segment and are therefore suitable for modeling similar physical problems. 
We discuss logopoles' properties, their relationships to spherical and spheroidal harmonics, and possible applications.
Many secondary proofs are given as supplementary material \cite{SI}. 
Our discussion of logopoles will be here restricted to harmonics with $m = 0$, but they can
be generalized to arbitrary $m$. This generalization, although
conceptually similar, is not straightforward and
brings added technicalities, so will be presented elsewhere
to allow us to focus here on the concepts rather than the
mathematics.
This work suggests that the SSHs of the second kind and the related logopoles may provide a fruitful alternative to
the common SSHs and PSSHs in some problems involving Laplace equation. We believe the greatest benefits of this new
approach could be realized when extending it to other related equations, notably the Helmholtz equation.

%
We first present a new formula expressing PSSHs as a finite sum of offset SSHs of the second kind at origins O' and O'':
\begin{align}
Q_n(\xi)&P_n(\eta)=\sum_{k=0}^n\frac{(n+k)!}{2k!^2(n-k)!(2R)^k}\nonumber\\
&\times\left[(-1)^{n+k}r''^kQ_k(\cos\theta'')-r'^kQ_k(\cos\theta')\right]. \label{QPvsSQpandSQppm} 
\end{align}
$P_n$ are the Legendre polynomials and $Q_n$ the Legendre functions of
the second kind.
Spherical coordinates $(r,\theta,\phi)$ (with $u\equiv\cos\theta$) are centered at the origin O, cylindrical coordinates are denoted $(z,\rho,\phi)$, and
the offset coordinates (see Fig.~\ref{CoordinateSystems}) can be expressed as
\begin{align*}
\begin{array}{lll}
\rho'&=\rho=\sqrt{x^2+y^2}\\
z'&=z-R\\
r'&=\sqrt{\rho^2+(z-R)^2}\\
u'&=\cos\theta' = z'/r'
\end{array}
\qquad
\begin{array}{lll}
\rho''&=\rho\\
z''&=z+R\\
r''&=\sqrt{\rho^2+(z+R)^2}\\
u''&=\cos\theta'' = z''/r''.
\end{array}
\end{align*}
The prolate spheroidal coordinates ($\xi$,$\eta$,$\phi$) are taken with foci at O' and O'' and defined as in \cite{Morse1953}:
\begin{align*}
\xi= \frac{r''+r'}{2R}, \qquad \eta=\frac{r''-r'}{2R}.
\end{align*}
The right hand-side of Eq.~\eqref{QPvsSQpandSQppm} is a generalization of the simple example in Eq.~\eqref{EqnIntro}. It is a carefully chosen combination of offset SSHs of the second kind that ensures their singularities at infinity cancel out. The resulting sum is only singular on the segment from O' to O'' and happens to correspond to the PSSHs.
It is known that $Q_n(\xi)P_n(\eta)$ can also be expanded as a series of spherical harmonics of the first kind \cite{jeffery1916relations}, but the expansion is infinite and only converges outside the sphere of radius $R$ centered at the origin. In contrast, the sum in Eq.~\eqref{QPvsSQpandSQppm} is finite and is valid everywhere except on the singularity line from O' to O''.
Eq.~\eqref{QPvsSQpandSQppm} raises the prospect of the more general applicability of SSHs of the second kind. Despite the relative simplicity of this expression, we could not find a simple proof for it, but
propose a more indirect proof, which is interesting in its own right as it leads us to introduce new functions: logopoles.

\begin{figure}
\includegraphics[scale=.74]{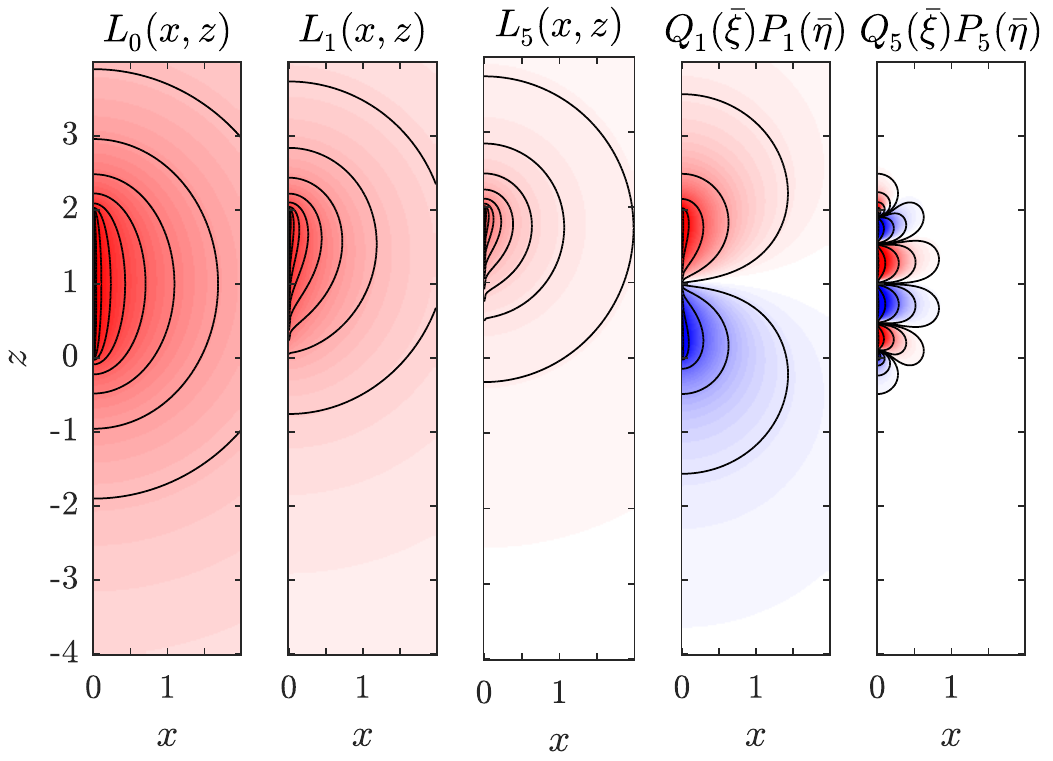}
\caption{Intensity plots of a selection of low order logopoles and  offset PSSHs with $R=2$. For better visualization, the functions have been rescaled and transformed by taking the arcsinh, which is similar to plotting on a log scale but allows for negative values. Red is positive, white is zero, and blue negative.
The solid lines represent equipotentials.}\label{logoplots}
\end{figure}

The logopoles can be formally defined as an infinite series of multipoles centered at O:
\begin{align}
L_n(\hat{r},\theta,\phi)&=\sum_{k=0}^{\infty} \frac{S_k}{n+k+1}\label{LvsS},
\end{align}
where $S_n(\hat{r},\theta,\phi)=\hat{r}^{-(n+1)}P_n(\cos\theta)$ are the external SSHs of the first kind.
For convenience, we have defined adimensional ``hat'' coordinates that are scaled by $R$, for example $\hat{r}=r/R$. 
$R$ is here an arbitrary length to make the functions adimensional and determines the scale of the logopoles.
It will become important when relating the logopoles to PSSHs.
This series diverges for $r< R$ but has the following analytic continuation to all space except the line segment $0\leq z\leq R$ on the $z$ axis:
\begin{align}
L_n= \tilde{S}_n - \sum_{k=0}^{n} \binom{n}{k}\tilde{S}_k^{\prime},
\label{LvsSQ}
\end{align}
where $\tilde{S}_n(\hat{r},\theta,\phi)=\hat{r}^n Q_n(\cos\theta)$ are the ``internal'' SSHs of the second kind
and the prime means that the function is of primed coordinates: $\tilde S_n^{\prime}=\tilde S_n(\hat{r}',\theta',\phi)$.
Eq.~\eqref{LvsSQ} is an alternative definition of the logopoles.
It provides the link with Eq.~\eqref{QPvsSQpandSQppm} and suggests 
that logopoles can be viewed as a regularization of the internal SSHs of the second kind ($\tilde{S}_n$).
The latter is singular of the $z$-axis and diverges at $\infty$, but the finite sum of offset SSHs removes
this singularity and divergence such that the logopoles are singular only on the $z$ axis between $0\le z \le R$.
Explicit expressions for the lowest orders of the logopoles (derived from Eq.~\eqref{L isolated} proved later) are given below:
\begin{align}
L_0=&\ln\frac{r'-z'}{r-z} = \ln\frac{r+z}{r'+z'} = Q_0(\xi) \\
L_1=&[uL_0-1]\hat{r} + \hat{r}'\\
L_2=&\frac{1}{2}[(3u^2-1)L_0-3u]\hat{r}^2  + \frac{1}{2}\left[3u\hat{r}+1\right]\hat{r}'
\end{align}
Fig.~\ref{logoplots} presents plots of low order logopoles and PSSHs. 

We now sketch a derivation of Eq.~\eqref{LvsSQ}, which gives further insight into the link between logopoles and multipoles.
We start from the well-known expansion of an offset point charge in terms of centered multipoles:
\begin{align}
S^{'}_0 &= \frac{R}{r'} =\sum_{k=0}^\infty~ S_k.  \label{offset chg}
\end{align}
and note that differentiation along $z$ is a ladder operator for SSHs \cite{1998NiceSummary}, explicitly:
\begin{align}
R\partial_z S_n&=-(n+1)S_{n+1}. \label{pdzS}
\end{align}
For this proof we will use integration instead of differentiation to move down the other way on the multipole ladder:
\begin{align}
&\int\d \hat{z} S_n =-\frac{S_{n-1}}{n} + f_1(\rho), \qquad(n\geq1)\label{Sn int}\\
&\int\d \hat z S_0 = \tilde{S}_0  + f_2(\rho), \label{S0 int} 
\end{align}
with $f_i(\rho)$ arbitrary functions. Eq.~\eqref{Sn int} derives from the ladder operator expression and Eq.~\eqref{S0 int} is
easily checked by integrating explicitly.
By integrating Eq.~\eqref{offset chg}, we then obtain
\begin{equation}
\tilde{S}_0^{'} = \tilde{S}_0 - \sum_{k=0}^\infty~ \frac{S_k}{k+1} +f(\rho) \label{EqnQoffset2}.
\end{equation}
We prove in Sec.~\refLvsSQproof~\cite{SI} that $f(\rho)=0$ and recognize the infinite series definition of $L_0$, which therefore satisfies $L_0=\tilde{S}_0-\tilde{S}_0^{'}$. Since the series in Eq.~\eqref{EqnQoffset2} converges for $z>R$, we must have that $L_0$ is also finite here, even though $\tilde{S}_0=Q_0(\cos\theta)$ and $\tilde{S}_0^{'}$ are singular on the axis ($\theta=0$). 
This proves Eq.~\eqref{LvsSQ} for $n=0$.
Logopoles of higher order $n$ can be obtained through repeated integration with respect to $z$, as shown in Sec.~\refLvsSQproof.

The ladder operator applies to logopoles in a similar way as to SSHs of the second kind:
\begin{align}
R\partial_z \tilde{S}_n&=n\tilde{S}_{n-1}, \label{pdzSQ}\\
R\pd_z L_n&=n~L_{n-1}-S_0^{\prime}. \label{L dif}
\end{align}
The latter is proved by applying the ladder operators to the series definition of $L_n$ (Eq.~\eqref{LvsS}) and using Eqs.~\eqref{offset chg} and \eqref{pdzS}.
This simple relation is in stark contrast with that for the spheroidal harmonics, where the operator results in an infinite series \cite{matcha1971prolate}:
\begin{align}
R\pd_z ~ Q_n(\xi)P_n(\eta)  = -2\sum_{\substack{k=n+1 \\ k+n\text{~odd}}}^\infty (2k+1)Q_k(\xi)P_k(\eta). \label{pdzQP}
\end{align}
We can also show (See Sec.~\refrecLproof) that the logopoles obey the following recurrence relation (for $n\ge 1$):
\begin{align}
(n+1)L_{n+1}-\hat{z}(2n+1)L_n +\hat{r}^2nL_{n-1} = \hat r', \label{Lrec}
\end{align}
which, up to the inhomogeneous term $\hat{r}'$, is identical to the recurrence for $\tilde S_n$, which can be derived from the recurrence for $Q_n(\cos\theta)$. Again, the PSSHs do not obey a similar simple recurrence. These recurrence properties Eqs. \eqref{L dif} and \eqref{Lrec}, show that logopoles are in some respect much closer to SSHs than to PSSHs despite having a line singularity like the latter.  
In addition, the proof by integration along with the radial dependences ($r^{n}$ vs $r^{-n-1}$) and Eq.~\eqref{pdzSQ} all suggest that the internal SSHs of the second kind can be viewed as the extension of external SSHs of the first kind to $n<0$. Although the link is not rigorous, we can write informally that $\tilde{S}_n\equiv S_{-n-1}/0$ where the division by zero represents what would result if we naively extrapolated Eq.~\eqref{Sn int} through $n=0$ \footnote{Technically $r^nP_n(\cos\theta)$ are the SSHs for $n<0$, since $P_{-n-1}=P_n$, but these do not fit on the same ladder described by the operator $\pd_z$.}. As $L_n$ provide a regularization of $\tilde{S}_n$ to ensure the singularity remains bounded, the logopoles can therefore be viewed as the most physical definition for multipoles of negative orders.




We now summarize useful additional properties of logopoles. While the expression of logopoles in terms of offset SSHs of the second kind, Eq.~\eqref{LvsSQ}, is the analytic continuation of the logopoles in all space, it is not obvious that the logopoles are finite on the $z$ axis for $z>R$ and $z<0$. To show this, we can express the Legendre functions as $Q_n=P_nQ_0-W_{n-1}$  
where
\begin{align}
W_{n-1}(x)=\sum_{k=1}^n\frac{P_{k-1}(x)P_{n-k}(x)}{k}
\end{align}
is a polynomial of degree $n-1$ \cite{Abramowitz}.
We then use the translation relation for internal spherical harmonics \cite{hobson1944theory}:
\begin{align}
\sum_{k=0}^n\binom{n}{k} \hat r'^kP_k(u')=\hat r^nP_n(u)
\end{align} 
to isolate the logarithmic part in Eq.~\ref{LvsSQ}:
\begin{align}
L_n=\hat r^n[P_n(u)L_0+W_{n-1}(u)] - \sum_{k=0}^n\binom{n}{k}\hat r'^kW_{k-1}(u'). \label{L isolated}
\end{align}
The singularity on $0<z<R$ is then entirely contained within $L_0$. 
As an alternative series definition, we can also express $L_n$ as a series of SSHs in the O' offset frame (see Sec.~\refsecLvsSp~for proof):
\begin{align}
L_n=\sum_{k=0}^\infty (-1)^{k}\frac{n!k!}{(n+k+1)!}S_k^{\prime},  \qquad (r'>R). \label{LvsSp}
\end{align} 
%



Having presented the basic properties of logopoles, we now derive the relations linking logopoles to spheroidal harmonics,
from which a proof of Eq.~\ref{QPvsSQpandSQppm} will result.
%
%
%
Since the singularity of the logopoles lies on the segment OO' and that of PSSHs on O'O'',
we first define a translated spheroidal coordinate system with foci at O and O' denoted $\xibar,\etabar$:
\begin{align}
\xibar=\frac{r+r'}{R},\quad \etabar=\frac{r-r'}{R}. \label{bar}
\end{align}
Then we can show that (see Sec.~\refproofQPvsL):
\begin{align}
Q_n(\xibar)&P_n(\etabar) = \sum_{p=0}^n\frac{(-1)^{p+n}(n+p)!}{2~p!^2(n-p)!} L_p. \label{QPbarvsL}
\end{align}
The inverse relationship (proved in Sec.~\refproofLvsQP) is
\begin{align}
L_n = \sum_{k=0}^n \frac{2n!^2(2k+1)}{(n-k)!(n+k+1)!} Q_k(\xibar)P_k(\etabar)\label{LvsQPbar}.
\end{align}
From these, we see that logopoles for $n\le N$ span the same space as offset PSSHs with $n\le N$, providing an alternative basis to that space.
We have not been able at this stage to prove or disprove the completeness of the infinite set of $L_n$.
We also note that these expansions have the same coefficients (up to a factor of 2) as those that relate internal PSSHs $P_n(\xibar)P_n(\etabar)$ to internal SSHs $r^nP_n(\cos\theta)$ (see Sec.~\refproofLvsQP), in which $r^nP_n(\cos\theta)$ would take the place of $L_n$, and $P_n(\xibar)$ the place of $Q_n(\xibar)$. This is yet another similarity between logopoles and multipoles.

We now return to the proof of Eq.~\ref{QPvsSQpandSQppm}. Substituting Eq.~\eqref{LvsSQ} into Eq.~\eqref{QPbarvsL} and simplifying using a binomial identity (see Sec.~\refproofQPbarMain), we
obtain the offset PSSHs as a sum of SSHs of the second kind:
\begin{align}
Q_n(\xibar)P_n(\etabar) = \sum_{k=0}^n\frac{(n+k)!}{2k!^2(n-k)!}\left[(-1)^{n+k}\tilde{S}_k - \tilde{S}_k^{\prime}\right].
\label{QPbarMain}
\end{align}
This relation is similar to Eq.~\ref{QPvsSQpandSQppm} except that the spheroidal coordinates are offset (singular on OO').
To find the equivalent relation for normal spheroidal coordinates (singular on O''O'), we apply the successive transformations
$R\rightarrow 2R$ and $z \rightarrow z+R$, which results in
Eq.~\ref{QPvsSQpandSQppm}, as required.

Finally, this link to PSSHs also allows us to investigate the source distributions that create these functions, or equivalently their integral forms. 
The spheroidal harmonics are known to be proportional to the potential of a charge distribution on a finite line segment, given by the Havelock formula \cite{havelock1952moment,miloh1974ultimate}:
\begin{align}
Q_n(\xi)P_n(\eta)=&R\int_{-1}^1 \frac{P_n(v)\d v }{\sqrt{\rho^2+(z-Rv)^2}}, \label{intQP}
\end{align}
or for offset PSSHs:
\begin{align}
Q_n(\xibar)P_n(\etabar)=&\frac{R}{2}\int_{0}^1 \frac{P_n(2v-1)\d v }{\sqrt{\rho^2+(z-Rv)^2}}. \label{intQPbar}
\end{align}
Substituting the latter in Eq.~\eqref{LvsQPbar}, we obtain after simplications (see Sec.~\refintLnmProof)
a similar integral form for logopoles: 
\begin{align}
L_n=&R\int_0^1\frac{ v^{n}\d v }{\sqrt{\rho^2+(z-Rv)^2}}.\label{intLnm}
\end{align}
While spheroidal harmonics are produced by Legendre polynomial charge distributions, logopoles are produced by monomial distributions. These also show that logopoles are positive functions and cannot therefore satisfy the same orthogonality relations as spherical harmonics.

It would be instructive to compare Eq.~\ref{intLnm} to an integral form for $\tilde{S}_n$, but none have been proposed to the best of our knowledge.
We show in Sec.~\refSQsourceproof~that $\tilde{S}_n$ may be expressed as a line source distribution on the entire $z$-axis. However, this source distribution diverges and the function must be expressed as the difference of the divergent line source and a sum of divergent multipoles from sources at $r=\infty$, both containing infinite charge. 
\begin{align}
\tilde{S}_n=&\lim_{a\rightarrow\infty} \Bigg\{\frac{R}{2}\int_{-a}^a\frac{\text{sign}(v)v^n\d v}{\sqrt{\rho^2+(z-Rv)^2}} \nonumber\\
& \hspace{2cm}-\sum_{\substack{k=0\\n-k\text{ odd}}}^{n}\frac{a^{n-k}}{n-k}\hat r^kP_k(\cos\theta). \Bigg\}\label{intSQ}
\end{align}
This new relation explains why these functions are neglected from physical analysis. Logopoles, in contrast, provide a regularization via truncation of the SSHs of the second kind, with an identical charge distributions on the $z$-axis but over a finite length.

\begin{figure}
\includegraphics[width=7.5cm, clip=true, ]{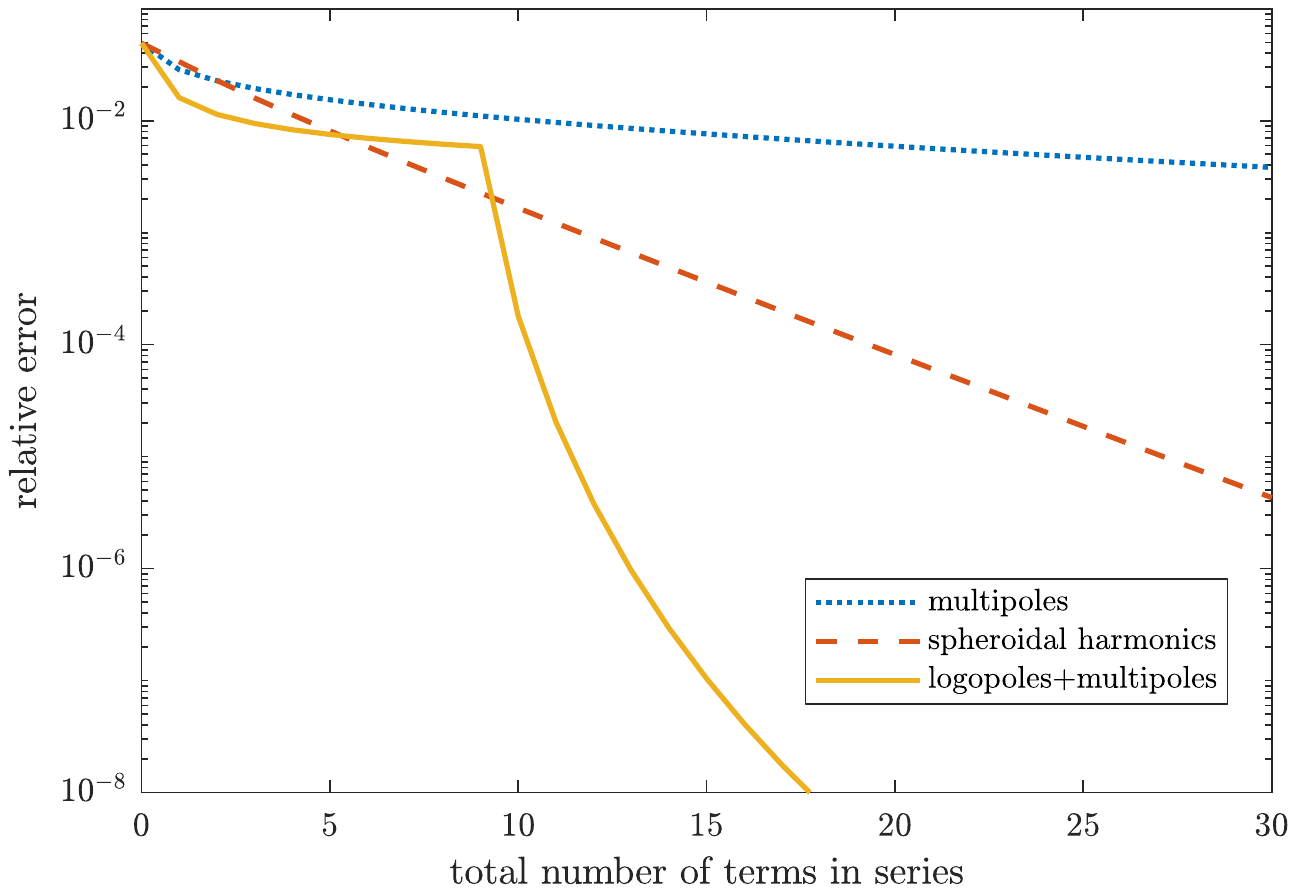}
\caption{Comparisons of the relative error in the computed reflected potential of a point charge near a dielectric sphere with dielectric constant 1.5.
The charge is located at a distance $0.01a$ from the sphere, with $a$ its radius, and the reflected potential is calculated at the point charge location using different series expansions and as a function of series truncation. The spherical and spheroidal harmonics solutions were discussed in detail in Ref.~\cite{majic2017super}.}
\label{logConvergence}
\end{figure}

We conclude by discussing possible implications of this work.
Logopoles are interesting in several respects. On the one hand they are strongly related to spheroidal harmonics through the finite sums Eqs.~(\ref{QPbarvsL}-\ref{LvsQPbar}) and sharing the same line singularity. One could then argue that there is no need for such new functions
since PSSHs could be used for anything where logopoles may be applicable. 
But on the other hand, in contrast to PSSHs, logopoles have a special link to spherical harmonics and can be viewed as regularized multipoles of negative order
with many similar properties. We believe this duplicity makes the logopoles a fruitful concept that deserves further investigation.
Also it highlights the fact that spherical harmonics of the second kind, which are most often neglected from physical analysis, can actually be used to construct \textit{localised} charge distributions, another concept worth additional investigation.
We have already generalized this work to the case of a general $m$ (this will be presented elsewhere). More work will be needed to assess whether similar functions can be related to the solid oblate spheroidal harmonics. More work will also be needed to improve the practical computation of logopoles, since the definitions given here are either numerically unstable in some regions of space for large $n$ or not computationally efficient.

In terms of applications, it is not yet clear where the largest advantages of these new concepts may lie, but we can point to an encouraging example: the potential of a point charge near a dielectric sphere. This classic problem can be solved using spherical harmonics
\cite{1941Stratton}, but the resulting series are very slowly converging when the point charge is close to the sphere \cite{moroz2011superconvergent}.
It was recently shown that using PSSHs instead of SSHs could dramatically improve the convergence of the solution \cite{majic2017super}.
We have found that logopoles form a solution that may converge even faster. The details of the derivation are outside the scope of this letter and will be presented elsewhere, but we nevertheless illustrate in Fig.~\ref{logConvergence} the improved convergence using a combination
of logopoles and multipoles.

Finally, we speculate that the greatest benefits of these new ideas may come from applying them to the Helmholtz equation $\nabla^2 V + k^2 V =0$.
The standard solutions in spheroidal coordinates, spheroidal wavefunctions, are not as user-friendly and well-behaved as the PSSHs, which render their application much more cumbersome \cite{1993Farafonov}. The ideas developed here could be applied to find better alternatives. For example Eqs.~\eqref{LvsS}, \eqref{intLnm}, and Eq.~A.5 in \cite{majic2017inside} can all be generalized to the scalar Helmholtz equation:
\begin{align}
H_n^{(1)}&=\sum_{k=0}^{\infty} \frac{h_k(kr)P_k(\cos\theta)}{n+k+1},\\
H_n^{(2)}&= R\int_0^1\frac{ v^{n}\d v ~e^{ik\sqrt{\rho^2+(z-Rv)^2}}}{\sqrt{\rho^2+(z-Rv)^2}},\\
H_n^{(3)}&= \sum_{k=n}^\infty \frac{k!^2~h_k(kr)P _k(\cos\theta)}{2(k-n)!(k+n+1)!} ,
\end{align}
with $h_k$ the spherical Hankel functions.
These are just a few possible alternatives to spheroidal wavefunctions, with identical line singularity in the long-wavelength limit.
These could provide a simpler or more efficient alternative for the solution of wave-scattering problems by spheroidal and elongated objects.
For all these reasons, we believe that logopoles and related functions will become a fundamental tool of mathematical physics, alongside multipoles and spheroidal harmonics.\\

\begin{acknowledgments}
ECLR acknowledges the support of the Royal Society Te Ap\=arangi (New Zealand) through a Marsden grant.
The authors are grateful to Baptiste Augui\'e and Dmitri Schebarchov for insightful discussions.
\end{acknowledgments}

	\newpage

	\renewcommand{\thesection}{S.\Roman{section}}
	
	\renewcommand{\thefigure}{S\arabic{figure}}
	\renewcommand{\thetable}{S.\Roman{table}}
	\renewcommand{\theequation}{S\arabic{equation}}

	\renewcommand\floatpagefraction{.99}
	\renewcommand\topfraction{.99}
	\renewcommand\bottomfraction{.99}
	\renewcommand\textfraction{.01} 
	
	\renewcommand{\thepage}{S\arabic{page}}

	\setcounter{equation}{0}
	\setcounter{page}{1}
	\textbf{~~~~~~~~SUPPLEMENTAL MATERIAL }
	\makeatletter
	\def\l@subsection#1#2{}
	\def\l@subsubsection#1#2{}
	\makeatother


	\section{Proof of Eq.~\refLvsSQ, analytic continuation of logopoles} \label{LvsSQproof}
	
	\subsection{Proof by induction}
	
	Starting from the series definition of logopoles Eq.~\refLvsS, we aim to prove their analytic continuation as a finite sum of offset spherical harmonics of the second kind Eq.~\refLvsSQ, explicitly:
	\begin{align}
	\tilde{S}_n - \sum_{p=0}^n\frac{n!\tilde{S}_p'}{p!(n-p)!} = \sum_{k=0}^\infty \frac{S_k}{k+n+1},  \label{trans SQ} 
	\end{align}
	which will be proved by induction on $n$. The base case was covered in the main text. Now assume Eq.~\ref{trans SQ} is valid for case $n$, we then integrate with respect to $z$ and reindex the summations to get up to an arbitrary function $f(\rho)$:
	\begin{align}
	&\frac{\tilde{S}_{n+1}}{n+1} - \sum_{p=0}^n\frac{n!\tilde{S}_{p+1}'}{(p+1)!(n-p)!} \nonumber\\
	&\qquad = \tilde{S}_0 - \sum_{k=1}^\infty \frac{S_{k-1}}{k(k+n+1)}   + f(\rho)\nonumber\\
	\Rightarrow& \tilde{S}_{n+1} -\sum_{p=1}^n\frac{(n+1)!\tilde{S}_p'}{p!(n+1-p)!} \nonumber\\
	&\qquad  = \tilde{S}_0 - \sum_{k=0}^\infty\bigg[ \frac{S_k}{k+1}-\frac{S_k}{k+n+2}\bigg] + f(\rho).\label{LnSQproof}
	\end{align}
	The sum of $S_k/(k+1)$ can be simplified using Eq.~\refEqnQoffset, and rearranging gives case $n+1$ up to some $f(\rho)$.

	\subsection{Proof of $f(\rho)=0$}
	
	First observe that $f(\rho)$ must be a solution of Laplace equation because it is a sum of other solutions.
	Laplace equation for $f(\rho)$ is simply $\partial_\rho [\rho \partial_\rho f]=0$, which has the general solution $f=f_0\ln(\rho/\rho_0)$. Then by showing that $f(0)=0$ it follows that $f(\rho)=0$. To show this we will evaluate Eq.~\ref{trans SQ} at $\rho=0$, $z>R$. The right hand side of Eq.~\ref{trans SQ} is
	\begin{align*}
	\mathrm{RHS}\ref{trans SQ}=\sum_{k=0}^\infty\frac{\hat r^{-k-1}}{n+k+1}.
	\end{align*}
	For the left hand side we use the equivalent Eq.~\refLisolated, which can be used on the $z$ axis without the problem of $\tilde{S}_n$ being singular on the entire axis. For $\rho=0,z>R$, we have $u=u'=1$, $\hat r'=\hat r-1$, and $W_{n-1}(1)=H_n=\sum_{k=1}^n 1/k$, the $n^{th}$ harmonic number. And $L_0$ is expressed using Eq.~\refLvsS~for $n=0$.  
	\begin{align*}
	&\mathrm{LHS}\ref{trans SQ}
	= \hat r^n[L_0 - H_n]+\sum_{k=0}^n\binom{n}{k}(\hat r-1)^kH_k\\
	&=\sum_{k=0}^\infty \frac{\hat r^{n-k-1}}{k+1} - \hat r^nH_n+\sum_{k=0}^n\sum_{p=0}^k\binom{k}{p}\binom{n}{k}H_k (-)^{k+p}\hat r^p\\
	&=\sum_{q=-n}^\infty\frac{\hat r^{-q-1}}{n+q+1} +\sum_{p=0}^{n-1}\sum_{k=p}^n\binom{n}{k}\binom{k}{p}H_k (-)^{k+p}\hat r^p 
	\end{align*}
	We then use the identity (proved below):
	\begin{align}
	\sum_{k=p}^n(-)^{k+p}\binom{n}{k}\binom{k}{p}H_k=\frac{-1}{n-p} \qquad (p<n) \label{HnID}
	\end{align}
	to obtain
	\begin{align*}
	\mathrm{LHS}\ref{trans SQ}
	&=\sum_{q=-n}^\infty\frac{\hat r^{-q-1}}{n+q+1} -\sum_{p=0}^{n-1}\frac{\hat r^p}{n-p} \nonumber\\
	&=\sum_{q=-n}^\infty\frac{\hat r^{-q-1}}{n+q+1} -\sum_{q=-n}^{-1}\frac{\hat r^{-q-1}}{n+q+1} =\mathrm{RHS}\ref{trans SQ}.
	\end{align*}
	This proves that $f(0)=0$ and therefore $f(\rho)=0$.
	
	\subsection{Proof of Auxiliary identity, Eq.\ref{HnID}}
	
	Eq.\ref{HnID} can be proved by induction on $p$. The base case for $p=0, n>0$ is Eq. 9.3a in \cite{2018bintrans} and is also proven in \cite{choi2011harmonicnumbers}. Assuming the identity is valid for a given $p$ and all $n>p$, we consider the $n+1$, $p+1$ cases. We will use the recurrence property of the harmonic numbers:  $H_{k+1}=H_k+1/(k+1)$. Then
	\begin{align}
	&\sum_{k=p+1}^{n+1}(-)^k\binom{n+1}{k}\binom{k}{p+1}H_k& \nonumber\\
	=&-\sum_{k=p}^{n}(-)^k\binom{n+1}{k+1}\binom{k+1}{p+1}\left[H_k+\frac{1}{k+1}\right] \nonumber\\
	=&-\frac{(-)^p}{n-p}\frac{n+1}{p+1} - \frac{1}{p+1}\sum_{k=p}^{n}(-)^k\binom{n+1}{k+1}\binom{k}{p},   \label{HnIDinduction}
	\end{align}
	where we used $\binom{k+1}{p+1}=\frac{k+1}{p+1}\binom{k}{p}$ and Eq.~\ref{HnID}. Now we derive another identity to simplify the sum over $k$, by applying the binomial theorem to $x^{n+1}$ twice:
	\begin{align}
	x^{n+1}&=\sum_{k=0}^{n+1}\binom{n+1}{k}(x-1)^k \nonumber\\
	&=1+\sum_{k=0}^n\binom{n+1}{k+1}(x-1)^k(x-1)  \nonumber\\
	&=1+\sum_{k=0}^n\binom{n+1}{k+1}\sum_{p=0}^k\binom{k}{p}(-)^{k+p}x^p(x-1). 
	\end{align}
	By rearranging the order of summation and matching the coefficients of each power of $x$, it must be that
	\begin{align}
	\sum_{k=p}^n\binom{n+1}{k+1}\binom{k}{p}(-)^{k+p}=1.
	\end{align}
	Insert this into Eq.~\ref{HnIDinduction} to show that the case $n+1$, $p+1$ holds. 
	
	\section{Proof of Eq.~\refLrec, recurrence relation for Logopoles} \label{recLproof}
	
	We start with Eq.~\refLvsSQ, the expression for logopoles in terms of offset spherical harmonics of the second kind:
	\begin{align}
	L_n= \tilde{S}_n-\sum_{k=0}^{n} \binom{n}{k}\tilde{S}_k^\prime,
	\end{align}
	and substitute it into the recurrence relation to show that the recurrence holds.
	The functions $\tilde{S}_n^{\prime}$ obey the same recurrence as logopoles without the inhomogeneous part $\hat{r}$. Then we must show that
	\begin{align}
	-(n+1)\sum_{k=0}^{n+1} {{n+1}\choose{k}}\tilde{S}_k^{\prime} +(2n+1)\hat{z}\sum_{k=0}^{n} {{n}\choose{k}}\tilde{S}_k^{\prime} \nonumber\\
	-n\hat{r}^2\sum_{k=0}^{n-1} {{n-1}\choose{k}}\tilde{S}_k^{\prime} =\hat{r}'.
	\end{align}
	This can be proved by writing $\hat{z}=\hat{r}'\cos\theta'+1$ and $\hat{r}^2=\hat{r}'^2 +2\hat{r}'\cos\theta'+1$, and equating the powers of $\hat{r}'^q$ for $0<q<n+1$, employing the recurrence relations for the binomial coefficients and Legendre functions of the second kind. 
	
	\section{Proof of Eq.~\refLvsSp, logopoles as a series of offset multipoles} \label{secLvsSp}
	
	The spherical solid harmonics can be expanded on an offset basis at O' \cite{hobson1944theory}:
	\begin{align}
	S_k=\sum_{p=k}^\infty(-)^{k+p}\binom{p}{k}S_p'.
	\label{EqnSvsSp}
	\end{align} 
	Inserting this into the original series definition of logopoles (Eq.~\refLvsS) and rearranging:
	\begin{align}
	L_n&=\sum_{p=0}^\infty  S'_p\sum_{k=0}^p \frac{(-)^{k+p}}{n+k+1}\binom{p}{k}
	&=\sum_{p=0}^\infty \frac{(-)^p n!p!}{(n+p+1)!}S_p'.
	\end{align}
	Here we used a binomial transform identity: Eq.~4.4 in Ref.~\cite{2018bintrans} for $y\equiv n+1$.

	\section{Proof of Eq.~\refQPbarvsL, expansion of offset spheroidal harmonics in terms of logopoles}
	\label{proofQPvsL}
	
	We first note the series expansions of the offset PSSHs $Q_n(\xibar)P_n(\etabar)$ in terms of SSHs
	(see for example Eq.~A.5 in Ref.\cite{majic2017inside}):
	\begin{align}
	Q_n(\xibar)P_n(\etabar)=&\frac{1}{2} \sum_{k=n}^\infty~\frac{k!^2}{(k-n)!(k+n+1)!} S_k.
	\end{align}
	Applying the transformation $z\rightarrow R-z$, we have $S_k \rightarrow (-)^k S'_k$, $r\rightarrow r'$ and $r'\rightarrow r$, so $Q_n(\xibar)\rightarrow Q_n(\xibar)$ and $P_n(\etabar)\rightarrow P_n(-\etabar) = (-)^n P_n(\etabar)$ giving:
	\begin{align}
	Q_n(\xibar)P_n(\etabar)=&\frac{1}{2} \sum_{k=n}^\infty~\frac{(-)^{k+n} k!^2}{(k-n)!(k+n+1)!} S'_k.
	\label{QPbarvsSbar}
	\end{align}
	
	Compare this to the series obtained by substituting Eq.~\refLvsSp~into the right-hand side of Eq.~\refQPbarvsL~and swapping the sums:
	\begin{align}
	&\sum_{p=0}^n\frac{(-)^{p+n}(n+p)!}{2p!^2!(n-p)!} L_p = \nonumber\\
	&\sum_{k=0}^\infty \frac{(-)^{k+n}}{2} S_k^{\prime}\sum_{p=0}^n \frac{(-)^{p}(n+p)!}{p!(n-p)!(p+k+1)!}.
	\end{align}
	This is the same as Eq.~\ref{QPbarvsSbar}, as required, thanks to the following identity:
	\begin{align}
	&\frac{k!}{(n-k)!(k+n+1)!}=\sum_{p=0}^n \frac{(-)^p(n+p)!}{(n-p)!p!(k+p+1)!} 
	\nonumber\\
	\Leftrightarrow& \sum_{p=0}^n(-)^p\binom{n+p}{n}\binom{k+n+1}{n-p}=\binom{k}{n}. \label{binom id}
	\end{align} 
	
	The last equality is found by writing: 
	\begin{align}
	&\frac{1}{(x+1)^{n+1}}(x+1)^{n+k+1}=(x+1)^k \\
	\Leftrightarrow \sum_{p=0}^\infty(-)^p&\binom{n+p}{n}x^p\sum_{t=0}^{n+k+1}\!\binom{n+k+1}{t}x^t=\sum_{q=0}^k\binom{k}{q}x^q 
	\end{align}
	and equating the coefficients of $x^n$. For each $p$ on the left hand side there will be terms containing $x^n$ for each $t=n-p$.
	
	\section{Proof of Eq.~\refLvsQPbar, expansion of logopoles in terms of offset spheroidal harmonics}
	\label{proofLvsQP}
	
	This proof is based on the recognition that the expansion of $Q_n(\xibar)P_n(\etabar)$ in terms of $L_n$
	(Eq.~\refQPbarvsL) exhibits the same expansion coefficients as that of $P_n(\xibar)P_n(\etabar)$ in terms
	of internal SSHs (Eq.~A.2 in Ref.~\cite{majic2017inside} for $m=0$), explicitly:
	\begin{align}
	P_n(\xibar)&P_n(\etabar) = \sum_{p=0}^n\frac{(-)^{p+n}(n+p)!}{p!^2(n-p)!} \hat{r}^p P_p(\cos\theta). \label{PPvsS}
	\end{align}
	The inverse relation is given in Eq.~A.4 in Ref.~\cite{majic2017inside}, and the expansion of $L_n$ in terms of $Q_p(\xibar)P_p(\etabar)$ must have the same expansion coefficients, due to the orthogonality property of the expansion coefficients themselves. Explicitly, if substituting Eq.~A.2 of Ref.~\cite{majic2017inside} into Eq.~A.4 of Ref.~\cite{majic2017inside}, setting $m=0$, swapping the order of the sums, and using the fact that the $S_n$'s form a basis, we obtain the combinatorial identity:
	\begin{align}
	\sum_{p=0}^n \sum_{k=p}^n \frac{(-)^{k+p}(2k+1)(k+p)!}{(n-k)!(n+k+1)!(k-p)!} = \delta_{n,p}.
	\label{DeltaId}
	\end{align}
	We then consider the quantity
	\begin{align}
	F_n=&\sum_{k=0}^n \frac{2(2k+1)n!^2}{(n-k)!(n+k+1)!}Q_k(\xibar)P_k(\etabar),
	\end{align}
	and insert the expansion of spheroidal harmonics in terms of $L_p$ (Eq.~\refQPbarvsL), and rearrange the order of summation:
	\begin{align}
	F_n=& \sum_{p=0}^n \frac{n!^2}{p!^2} \sum_{k=p}^n \frac{(-)^{k+p}(2k+1)(k+p)!}{(n-k)!(n+k+1)!(k-p)!} L_p = L_n,
	\end{align}
	which proves Eq.~\refLvsQPbar.

	\section{Proof of Eq.~\refQPbarMain, expansion of offset spheroidal harmonics in terms of spherical harmonics of second kind}
	\label{proofQPbarMain}
	
	We first substitute Eq.~\refLvsSQ~into Eq.~\refQPbarvsL:
	\begin{align}
	Q_n(\xibar)P_n(\etabar) &= \sum_{k=0}^n\frac{(n+k)!}{2k!^2(n-k)!}\nonumber\\
	&\times\left[(-)^{n+k}\tilde{S}_k -
	\sum_{p=0}^k (-)^{n+k} \binom{k}{p}\tilde{S}_p^{\prime}\right].
	\label{QPforDoubleSum}
	\end{align}
	The double sum can be simplified by swapping the summation order to:
	\begin{align}
	\Sigma\Sigma = \sum_{p=0}^n \tilde{S}_p^{\prime} \sum_{k=p}^n (-)^{n+k} \binom{k}{p} \frac{(n+k)!}{2k!^2(n-k)!}.
	\end{align}
	Using the following identity which we will prove below:
	\begin{align}
	\sum_{k=p}^n\frac{(-)^k(n+k)!}{k!(k-p)!(n-k)!}=\frac{(-)^n(n+p)!}{p!(n-p)!},
	\label{BinId2}
	\end{align}
	we deduce
	\begin{align}
	\Sigma\Sigma = \sum_{p=0}^n \frac{(n+p)!}{2p!^2(n-p)!} \tilde{S}_p^{\prime} .
	\end{align}
	Substituting back into Eq.~\ref{QPforDoubleSum}, and re-indexing $k\rightarrow p$, we obtain Eq.~\refQPbarMain~as required.
	
	In order to prove the combinatorial identity (Eq.~\ref{BinId2}),
	we start from the expansions of $P_n(x)$ in powers of $(x+1)$ and $(x-1)$:
	\begin{align}
	P_n(x)&=\sum_{p=0}^n\frac{(-)^{n+p}(n+p)!}{p!^2(n-p)!} \frac{(x+1)^p}{2^p} \label{BinId2a}\\
	&=\sum_{k=0}^n \frac{(n+k)!}{k!^2(n-k)!}\frac{(x-1)^k}{2^k}.\label{BinId2a2}
	\end{align}
	Then expressing $[(x-1)/2]^k$ as a binomial series of $(x+1)/2$:
	\begin{align}
	P_n(x)=\sum_{k=0}^n\frac{(n+k)!}{k!^2(n-k)!}\sum_{p=0}^k(-)^{k+p}\binom{k}{p}\frac{(x+1)^p}{2^p}. \label{BinId2b}
	\end{align}
	Rearranging the summation order, then equating each coefficient of $(x+1)^p$ in Eqs.~\ref{BinId2a} and \ref{BinId2b}
	gives the required identity Eq.~\ref{BinId2}.
	
	\section{Proof of Eq.~\refintLnm, integral form of logopoles}
	\label{intLnmProof}
	Substitute the integral form of PSSHs (Eq.~\refintQPbar), into the expansion of logopoles in terms of PSSHs (Eq.~\refLvsQPbar) to get
	\begin{align}
	L_n =& \frac{R}{2}\int_{0}^1 \frac{\d v} {\sqrt{\rho^2+(z-Rv)^2}}\nonumber\\
	&\times\sum_{k=0}^n \frac{2(2k+1)n!^2}{(n-k)!(n+k+1)!} P_k(2v-1).
	\end{align}
	This reduces to Eq.~\refintLnm~thanks to the following expansion
	\begin{align}
	v^n = \sum_{k=0}^n \frac{(2k+1) n!^2}{(n-k)!(n+k+1)!} P_k(2v-1),
	\label{BinId2a3}
	\end{align}
	which is the inverse of Eq.~\ref{BinId2a2} with $x=2v-1$, and derives directly from Eq.~\ref{DeltaId}.
	
	\section{Proof of Eq.~\refintSQ, line integral form for spherical harmonics of the second kind}
	\label{SQ source proof}
	
	First of all, informally the charge distribution $\text{sign}(v)v^n$ can be obtained from the behavior $r^n$ near the $z$-axis and the antisymmetry of $Q_n(\cos\theta)$ about $z$. But we will prove Eq.~\refintSQ~more formally by recurrence using:
	\begin{align}
	(n+1)\tilde{S}_{n+1}=(2n+1)\hat r u \tilde{S}_n - n\hat r^2 \tilde{S}_{n-1}.
	\end{align}
	The base cases $n=0,1$ can be obtained from direct evaluation of the integral, by splitting the integration as $\int_{-a}^a\text{sign}(v)=\int_0^a-\int_{-a}^0$. Now substituting the assumed integrals for $\tilde{S}_n,\tilde{S}_{n-1}$: 
	\begin{align}
	(n+1)\tilde{S}_{n+1}=
	&\lim_{a\rightarrow\infty}\Bigg\{\frac{2n+1}{2}\hat r u\int_{-a}^a\frac{\text{sign}(v)v^n\d v}{\sqrt{\hat r^2-2\hat r u v + v^2}} \nonumber\\&- \frac{n}{2}\hat r^2\int_{-a}^a\frac{\text{sign}(v)v^{n-1}\d v}{\sqrt{\hat r^2-2\hat r u v + v^2}} \nonumber\\
	&-\sum_{\substack{k=0\\n-k\text{ odd}}}^{n-1}\frac{a^{n-k}}{n-k}\hat r^{k+1}(2n+1)uP_{k}\nonumber\\
	&+\sum_{\substack{k=0\\n-k\text{ even}}}^{n-2}\frac{a^{n-1-k}}{n-1-k}\hat r^{k+1}nP_k\Bigg\}.
	\end{align}
	We will show that the right hand side leads to {Eq.~\refintSQ}
	for $\tilde{S}_{n+1}$. The two integrals can be dealt with identity 2.263.1 of \cite{tables2014} with $n\rightarrow 0, m\rightarrow n$, and re-indexing $k\rightarrow k-1$ in the sum for $\tilde{S}_{n-1}$. This gives
	\begin{align}
	&(n+1)\tilde{S}_{n+1}=
	\lim_{a\rightarrow\infty}\Bigg\{ \frac{n+1}{2}\int_{-a}^a\frac{\text{sign}(v)v^{n+1}\d v}{\sqrt{\hat r^2-2\hat{r}uv+v^2}} \nonumber\\
	&-\frac{1}{2}\big[\sqrt{\hat r^2-2\hat ruv+v^2}v^n\big]_{-a}^a\nonumber\\
	&-\sum_{\substack{k=0\\n-k\text{ odd}}}^{n-1}\frac{a^{n-k}}{n-k}\hat r^{k+1}\big[(2n+1)uP_{k}-nP_{k-1}\big]\Bigg\}. \label{n-1}
	\end{align}
	In this proof for convenience we use the convention $P_{n}=0$ if $n<0$.
	$\sqrt{\hat r^2-2\hat ruv+v^2}$ can be expanded as a series by integrating the generating function for the Legendre polynomials.
	\begin{align}
	\sqrt{\hat r^2-2\hat r uv+v^2}=&\sum_{k=-1}^\infty\frac{\hat{r}^{k+1}}{v^{k}}\frac{P_{k-1}-P_{k+1}}{2k+1}. \label{exp}
	\end{align}
	Substituting in the bounds and noting that in the limit $a\rightarrow\infty$, we can ignore terms in Eq.~\ref{exp} that lead to negative powers of $a$:
	\begin{align*}
	\frac{1}{2}&\big[\sqrt{\hat r^2-2\hat ruv+v^2}v^n\big]_{-a}^a \!\!=\!\!\!
	\sum_{\substack{{k=-1}\\n-k\text{ odd}}}^{n-1}\!\!a^{n-k}\hat r^{k+1}\frac{P_{k-1}-P_{k+1}}{2k+1}.
	\end{align*}
	Substituting this in Eq.~\ref{n-1}, using the recurrence relation for the Legendre polynomials and rearranging gives the required expression for $(n+1)\tilde{S}_{n+1}$.

	\bibliography{libraryH}

\end{document}